\documentclass[twocolumn,prl]{revtex4}
\usepackage{amsmath}
\usepackage{epsfig}
\newcommand{\mtin}[1]{\mbox{\tiny {#1}}}
\newcommand{\ca}[1]{{\cal #1}}
\newcommand{\sfrac}[2]{{\textstyle\frac{#1}{#2}}}

\begin{document}
\bibliographystyle{apsrev}
\title{Efficient Fermionic One--Loop RG for the 2D Hubbard Model at Van Hove Filling}
\author{C. Husemann}\email{c.husemann@thphys.uni-heidelberg.de}
\author{M. Salmhofer}\email{m.salmhofer@thphys.uni-heidelberg.de}
\affiliation{Institut f\"ur theoretische Physik, Universit\"at Heidelberg}
\date{\today}

\begin{abstract}
We propose a novel parametrization of the four--point vertex function
in the one--loop one--particle irreducible renormalization group (RG)
scheme for fermions. 
It is based on a decomposition of the effective two--fermion
interaction into fermion bilinears that interact via exchange
bosons.
Besides being more efficient than
previous $N$--patch schemes, this parametrization also reduces the
ambiguity of introducing boson fields. We apply this parametrization to the two--dimensional $(t,t')$--Hubbard model using a novel $\Omega$--frequency regularization.
\end{abstract}
\maketitle

{\em Introduction.} In the past decade the one--loop Wilsonian renormalization group (RG) has been extensively used to
study the weak coupling instabilities of the two--dimensional
$(t,t')$--Hubbard model \cite{ZanchiSchulz1,
ZanchiSchulz2000,HalbothMetzner,HalbothMetznerPomeranchuk,ZanchiSelfEnergy, 
  Umklapp,MetznerSelfEnergy,NpatchRG,
  TemperatureFlow,katanin2patch,KataninKampfSelfEnergy,gflow,
  HonerkampSelfEnergy}.
Successes include the explanation of $d$--wave pairing tendencies
from a repulsive interaction. The interplay of antiferromagnetic and
$d$--wave superconducting instabilities found for small next to
nearest neighbor hopping $-t'$ resembles the behavior of the high
temperature cuprates near half filling. Avoiding the artificial suppression of small momentum particle--hole fluctuations by regularizing 
with temperature, the leading instability for strong hopping $-t'$
was found to be ferromagnetism at Van Hove filling and triplet
superconductivity away from Van Hove filling \cite{TemperatureFlow,
TemperatureFlow2}.

The one--loop truncation in the one--particle irreducible RG scheme
results in an integro--differential equation for the four--point vertex
function and the self--energy \cite{salmfrgtt}. The studies mentioned in the last paragraph neglect the self--energy and the frequency dependence of
the four--point function. The remaining three independent momenta of
the four--point function (one is determined by momentum conservation) are numerically discretized with $N$
patches in momentum space. In most studies these have been chosen along the Fermi surface by power counting arguments. Then a system of ordinary
differential equations of size $\sim N^3$ has to be solved
numerically. Generically for the
two--dimensional Hubbard model at low temperature, some coupling constants, which describe the vertex function for certain momentum combinations, grow large. In order to
determine the nature of the corresponding instability,
susceptibilities are calculated in the flow by introducing external
boson fields coupled to appropriate fermion bilinears \cite{salmfrgtt}.

In this work we develop a novel parametrization of the four--point
function in the one--particle irreducible RG scheme.
Guided by the singular momentum structure of the right hand side of
the flow equation, we identify three channels with distinct singular
momentum structures. In these channels the effective two--fermion
interaction is expanded in fermion bilinears that interact via
exchange bosons. These bosons, however, are dealt with in a purely fermionic language. We show that only a small number of terms is
needed to capture the essential features of the one--loop RG flow.
This improves the efficiency of the parametrization compared to
previous $N$--patch schemes. A similar reduction has recently been used for the single impurity Anderson model \cite{Karrasch}. We apply the method to the
two--dimensional Hubbard model at Van Hove filling and compare the
results with the temperature RG flow \cite{TemperatureFlow}. Instead of regularizing with temperature, we introduce a novel frequency regularization scheme with scale parameter $\Omega$.

{\em Decomposition of the Effective Interaction.}
Although the initial vertex function of the Hubbard model at high scales is a
constant in momentum space, according to the RG equation a non--trivial momentum structure evolves
in the flow as the scale is lowered. The instabilities of the RG flow, are however mainly determined by the singular momentum structure of the right hand side of the RG equation. Therefore, we develop a parametrization of the vertex function that simplifies the momentum dependence but keeps track of all possible singular contributions. As long as the vertex function is
still regular, the only momentum dependence that can change the
singular behavior of the RG equation is the transfer momentum that
flows through the scale--derivative of the particle--particle
or particle--hole two--fermion bubble. 

\begin{figure}[htb]
\begin{tabular}{cl}
\begin{minipage}{3cm}
$\dot{\Phi}_{\mtin{SC}}^{\Omega}(k_1,k_3,k_1+k_2)$ 
\end{minipage}
&=$\;-$
\begin{minipage}{2cm}
\includegraphics[width=2cm]{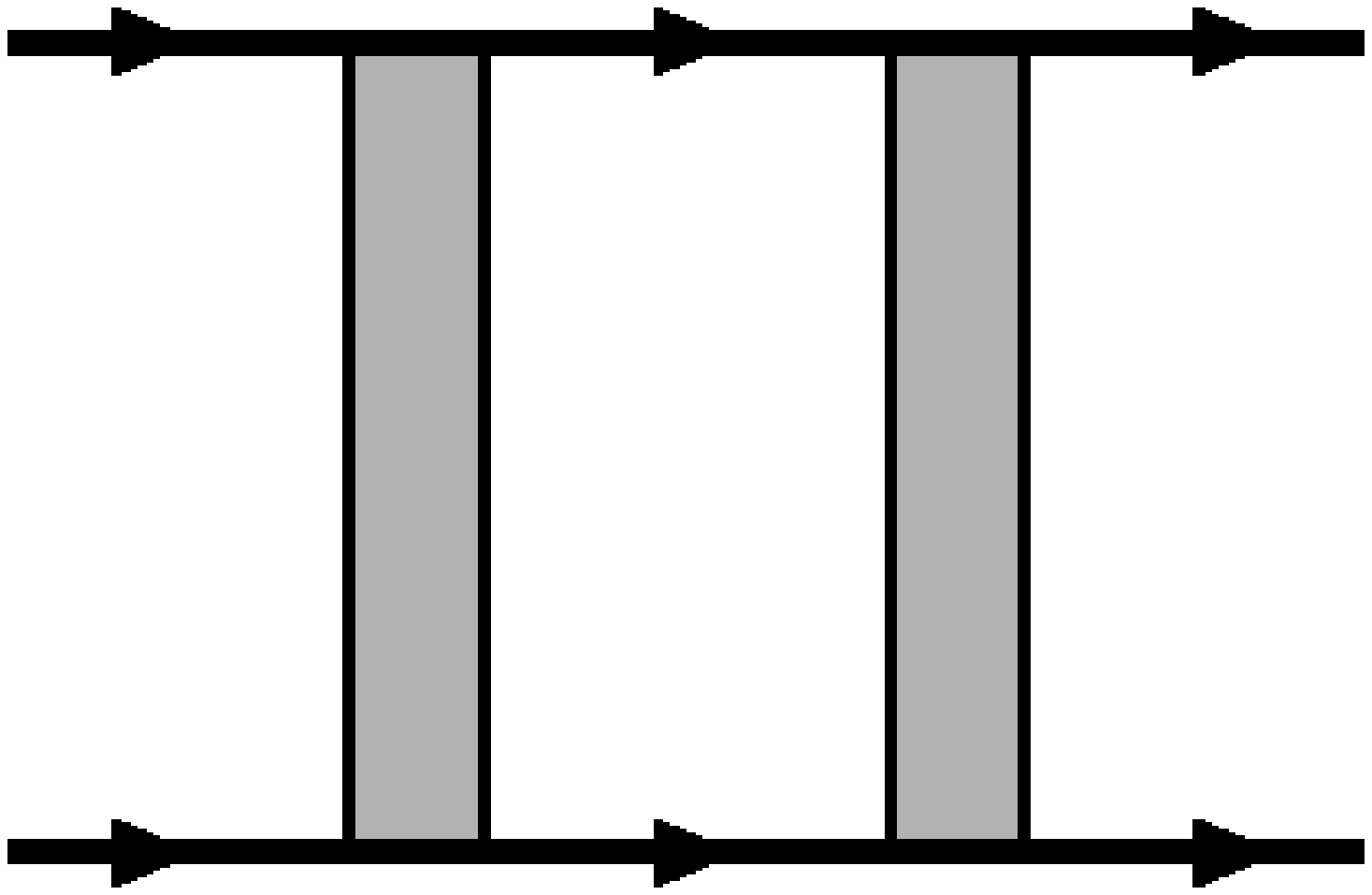}
\end{minipage} \\
\begin{minipage}{3cm}
$\dot{\Phi}_{\mtin{M}}^{\Omega}(k_1,k_2,k_3-k_1)$ 
\end{minipage}
&=
\begin{minipage}{2cm}
\includegraphics[width=2cm]{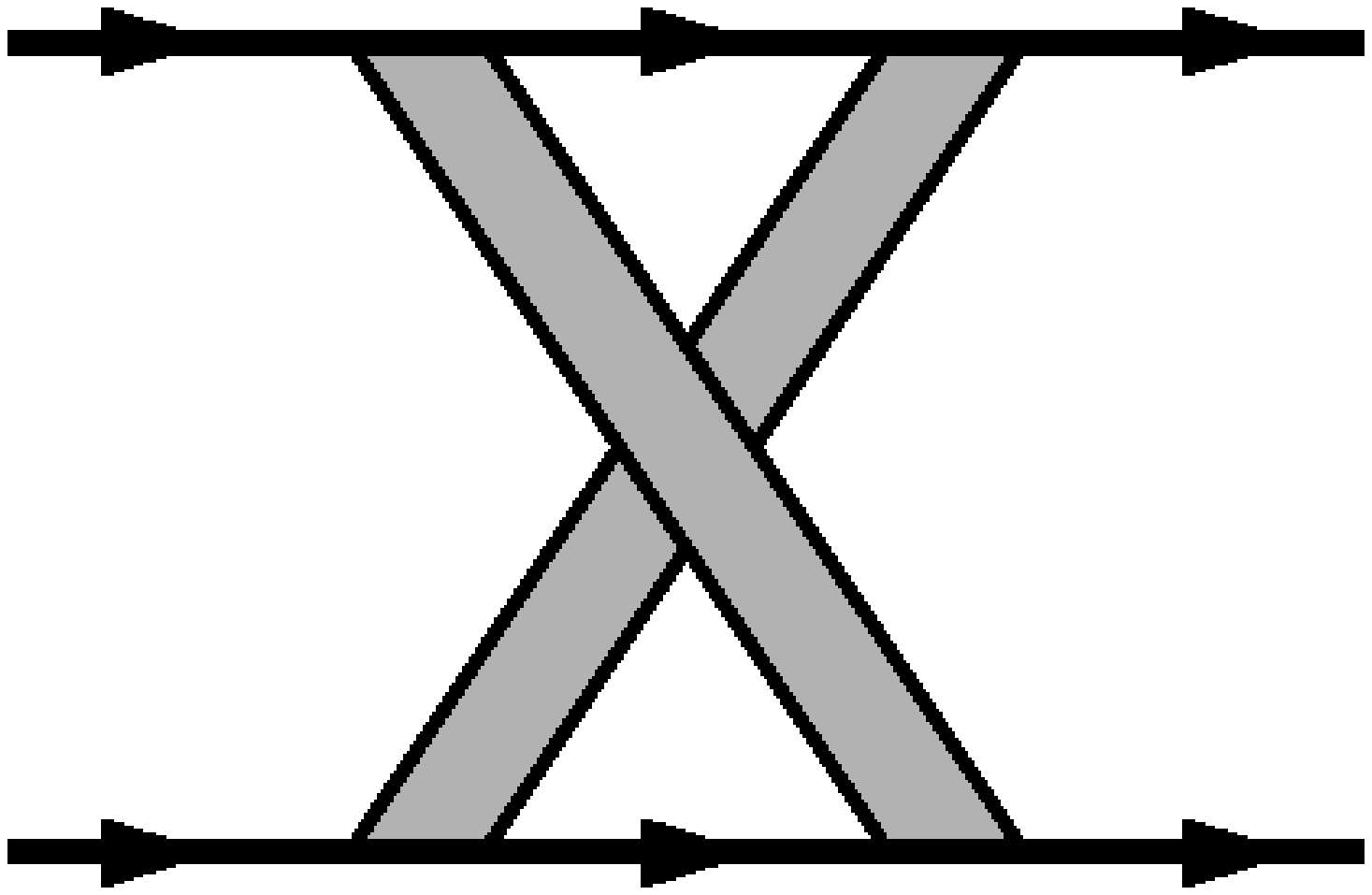}
\end{minipage} \\
\begin{minipage}{3cm}
$\dot{\Phi}_{\mtin{K}}^{\Omega}(k_1,k_2,k_2-k_3)$ 
\end{minipage}
&= 4
\begin{minipage}{1cm}
\includegraphics[width=1cm]{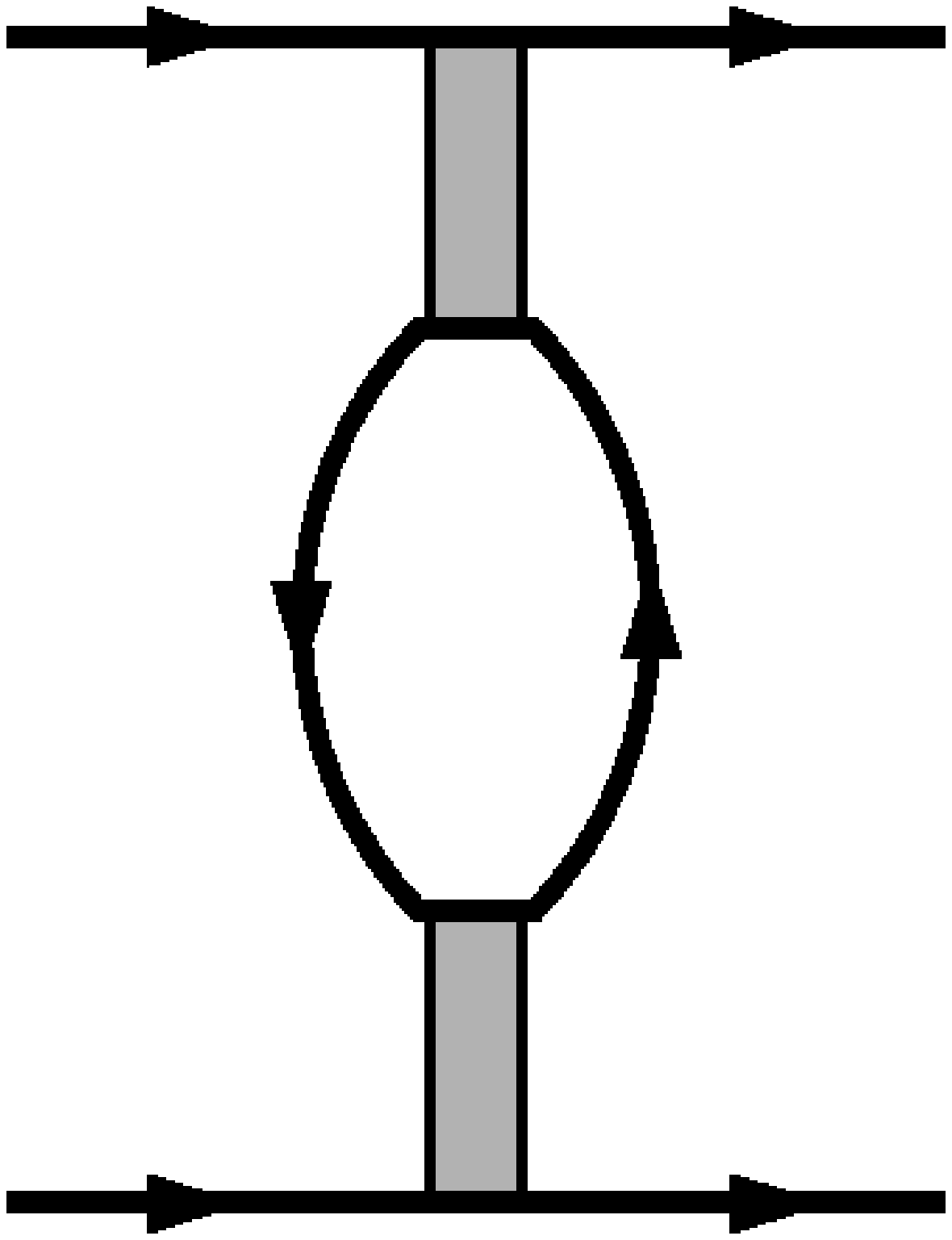}
\end{minipage}
$-2$ 
\begin{minipage}{1cm}
\includegraphics[width=1cm]{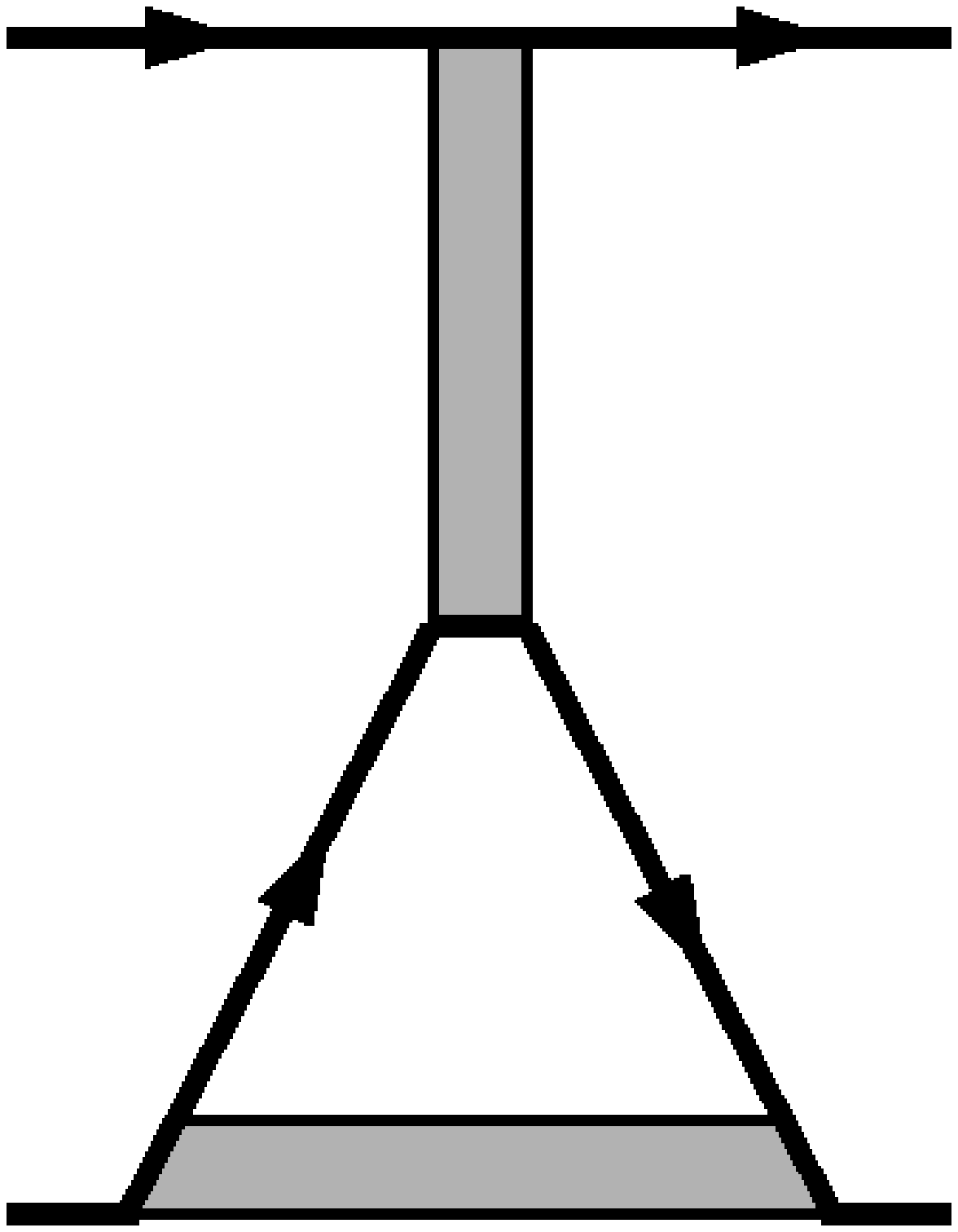}
\end{minipage} \\
\begin{minipage}{3cm}
\end{minipage}
&$\quad -2$ 
\begin{minipage}{1cm}
\includegraphics[width=1cm]{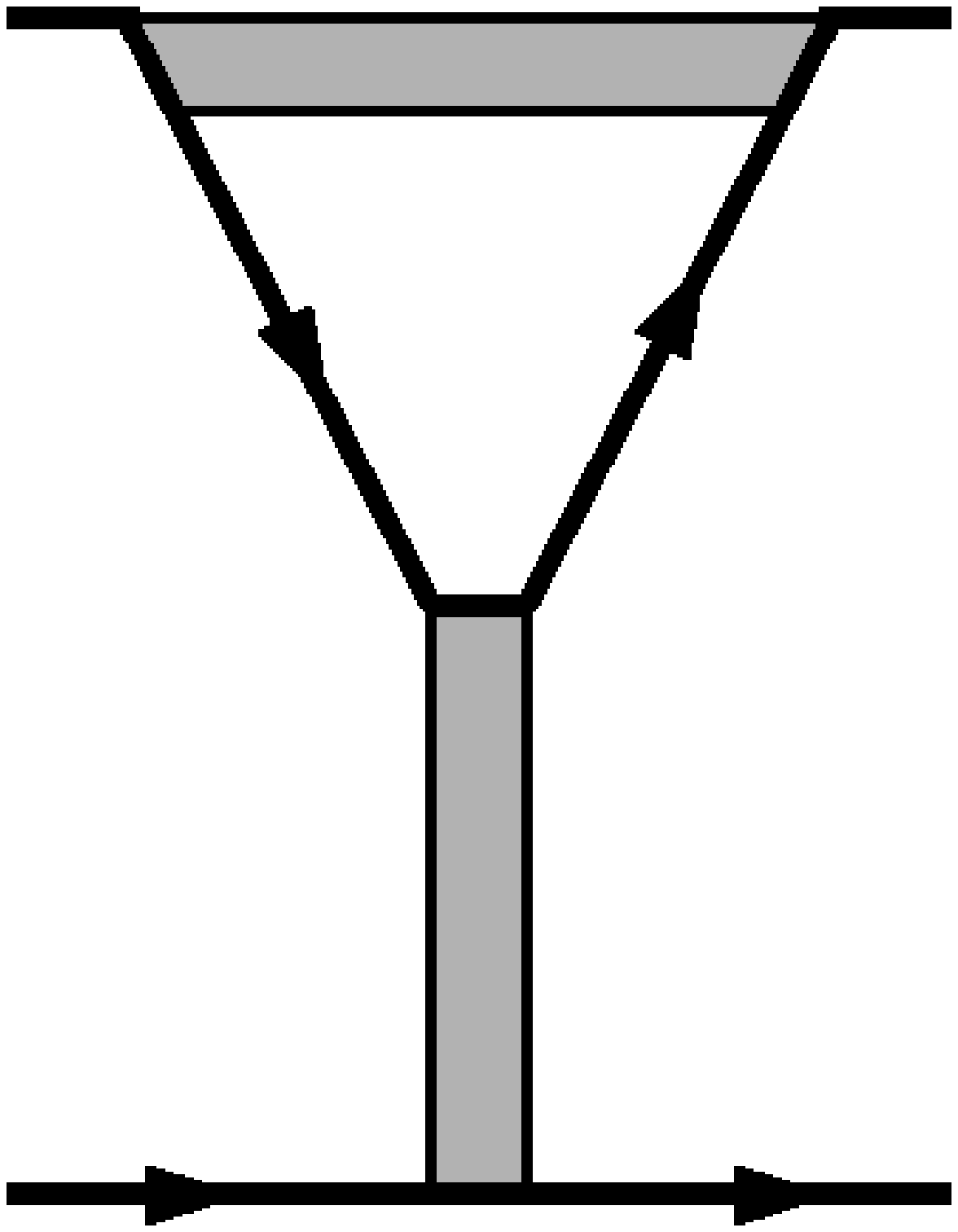}
\end{minipage}
+ 
\begin{minipage}{2cm}
\includegraphics[width=2cm]{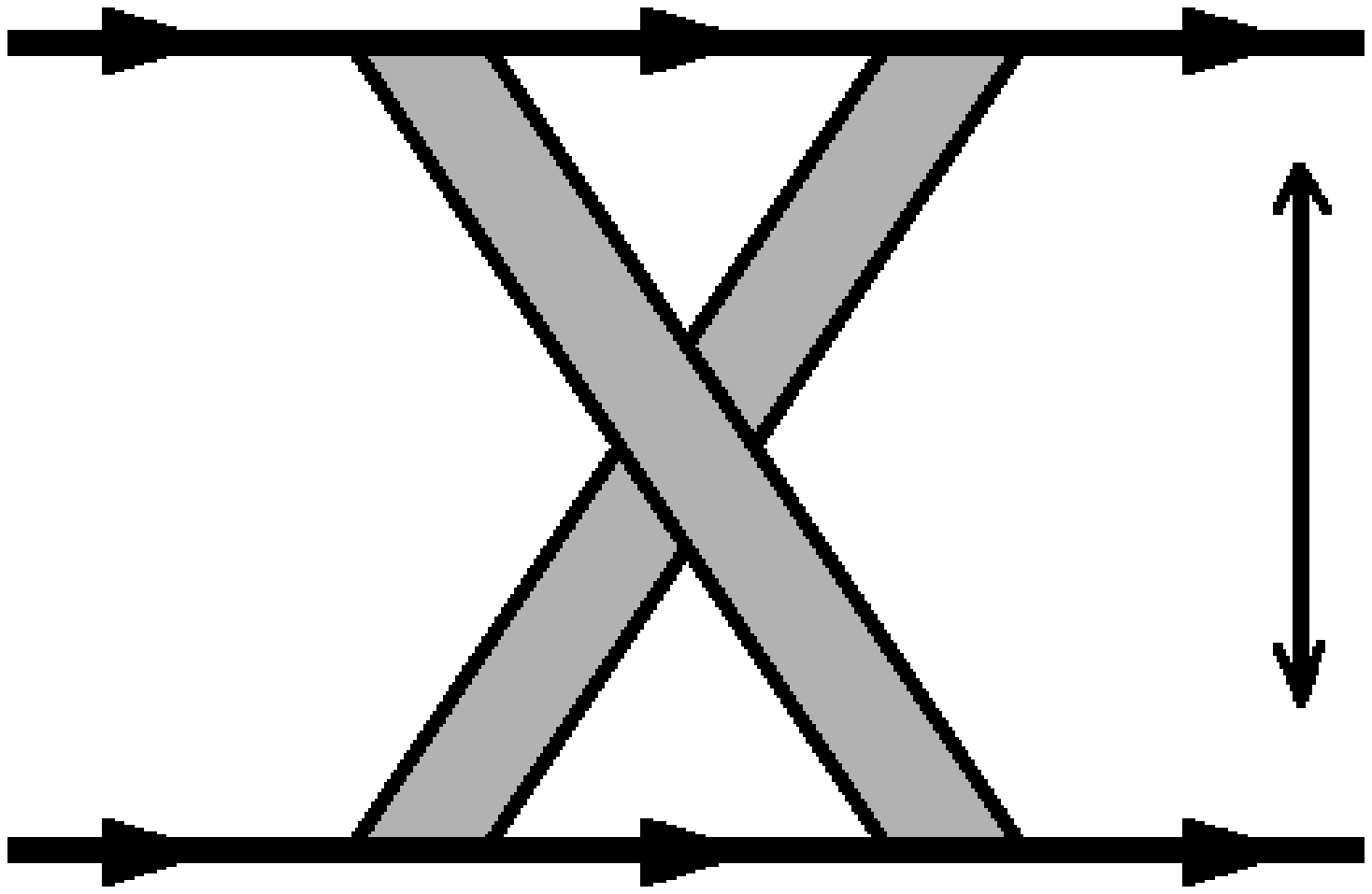}
\end{minipage}
\end{tabular}
\caption{Evolution of the superconducting, magnetic, and forward scattering channel respectively. The dot stands for a derivative with respect to the scale $\Omega$ and the doublesided arrow in the last graph symbols the exchange of two momenta and frequencies. \label{fig:channeldefinition}}
\end{figure}

On the right hand side of the RG equation three different classes of
graphs with distinct transfer momenta contribute -- the
particle--particle graph, the crossed particle--hole graph, and the
direct particle--hole graphs \cite{salmfrgtt}. Accordingly, we identify three
different channels, which we first assume as general charge and spin
rotation invariant two--fermion interactions. 
They are defined by three conditions: (1) each vertex function of the channels absorbs
one singular momentum, that is, one transfer momentum; (2) the channels describe the interaction of Cooper pairs, spin operators, and density operators, respectively, if the transfer momentum becomes singular; and (3)
each channel separately satisfies particle--hole symmetry and the fermionic antisymmetry.
Then the evolution of the
superconducting channel is given by the particle--particle graph, of
the magnetic channel by the crossed particle--hole graph, and of the
forward scattering channel by a combination of crossed and direct
particle--hole graphs, see Figure \ref{fig:channeldefinition}. The vertices in Figure \ref{fig:channeldefinition} depict the fermionic irreducible four--point vertex $V_{\Omega}(k_1,k_2,k_3)$, where we denote momentum $\mathbf{k}$ and frequency $k_0$ together by $k=(\mathbf{k},k_0)$. Spin is conserved along the fermion lines, see \cite{salmfrgtt}.

The initial condition for the channels is zero, so they are generated in the flow. 
The full vertex function of the Hubbard model is then given by the sum  
\begin{align}\label{eq:decompositionvertexfunction}\nonumber
&V_{\Omega}(k_1,k_2,k_3) = U 
-\Phi_{\mtin{SC}}^{\Omega}(k_1,k_3,k_1+k_2) \\& \quad +
\Phi_{\mtin{M}}^{\Omega}(k_1,k_2,k_3-k_1) + \frac 12
\Phi_{\mtin{M}}^{\Omega}(k_1,k_2,k_2-k_3) \\ &\quad- \frac 12
\Phi_{\mtin{K}}^{\Omega}(k_1,k_2,k_2-k_3) \, .\nonumber
\end{align}
This decomposition is exact on one--loop level since all graphs with their full momentum and frequency
dependence are taken into account.  
The assignment of the graphs, however, can be ambiguous for special
momenta. Using the Fierz identity, for example, a constant term can
be freely distributed among the channels. This is why the initial
Hubbard repulsion $U$ is not decomposed and kept explicitly in the
decomposition as a constant. However, this does not mean that the on--site term in
the full interaction remains independent of scale. The corrections
to it are rather absorbed as contributions to the above defined
three channels.

Next we further decompose the two--fermion interaction by writing
each channel as a sum of terms, in which two fermion bilinears interact via an
exchange boson propagator, where the boson momentum is the
corresponding singular transfer momentum of the channel. 
For fixed transfer momentum $l$ the vertex function
$\Phi_{\mtin{SC}}^{\Omega}(q,q',l)=\Phi_{\mtin{SC}}^{\Omega}(q',q,l)$ can be seen as the kernel of  a Fredholm operator. Therefore, due to
compactness, a diagonal expansion in eigenfunctions is possible in
principle. To avoid a calculation of the (scale--dependent) eigenfunctions, we simply expand in an orthonormal basis of $L^2((-\pi,\pi]^2)$ of form factors $f_n$ that represent the point symmetry group of the lattice. Then, the expansion involves non--diagonal terms 
\begin{align}\label{eq:SCexpansion}
\Phi_{\mtin{SC}}^{\Omega}(q,q',l) &= \sum_{m,n\in
\ca{I}} D_{mn}^{\Omega}(l) f_m(\sfrac
 {\mathbf{l}}{2}- \mathbf{q}) f_{n}(\sfrac{\mathbf{l}}{2}-\mathbf{q}') \nonumber \\  &\hspace{2cm}
 +R_{\mtin{SC}}^{\Omega}(q,q',l) \, .
\end{align} 
For simplicity the form functions are chosen
frequency independent here. The expansion coefficients $D_{mn}^{\Omega}(l)$
depend on the transfer momentum and frequency $l$ and are called superconducting boson propagators.
In practise, we restrict to a finite set $\ca{I}$ of terms. The
remainder function $R_{\mbox{\tiny SC}}^{\Omega}$ accounts for the
error.

Similarly, the magnetic and forward scattering channels are expanded
in appropriate frequency and scale independent form factors  
\begin{align}\label{eq:MKexpansion}
\Phi_{\mtin{M}}^{\Omega}(q,q',l)&=\sum_{m,n\in\ca{I}}
M_{mn}^{\Omega}(l) f_m(\mathbf{q}+ \sfrac{\mathbf{l}}{2}) \nonumber
f_n(\mathbf{q'}-\sfrac{\mathbf{l}}{2}) \\ &\hspace{2cm}+ R_{\mtin{M}}^{\Omega}(q,q',l)\\
\nonumber
\Phi_{\mtin{K}}^{\Omega}(q,q',l)&=\sum_{m,n\in\ca{I}}
K_{mn}^{\Omega}(l) f_m(\mathbf{q}+ \sfrac{\mathbf{l}}{2})
f_n(\mathbf{q'}-\sfrac{\mathbf{l}}{2}) \\ & \hspace{2cm} + R_{\mtin{K}}^{\Omega}(q,q',l) \, .
\end{align} 
This introduces the magnetic boson propagators $M_{mn}^{\Omega}$, the forward scattering boson propagators $K_{mn}^{\Omega}$, and additonal remainder terms.

{\em The Flow of Boson Propagators.}
The flow equations for the boson propagators are obtained by
inserting the proposed decomposition in the RG equation for the vertex function \cite{salmfrgtt} and
projecting each channel onto the coefficients of the orthogonal form
factor expansion. If the remainder terms are dropped, this gives a
closed system of integro--differential equations for the boson
propagators. Compared to the original RG equation for the vertex
function, their solution is now given by several functions dependent
on one momentum and frequency instead of one function dependent on
three momenta and frequencies. This is a considerable simplification 
and, in particular, makes the numerical implementation more
favorable. Furthermore, like the two--fermion bubbles, the boson propagators have only point singularities in momentum space instead of extended Fermi surface singularities. 

In the case of a regular Fermi surface with irrelevant Umklapp
scattering the RG flow is dominated by
the superconducting boson propagators if the initial
interaction contains an attractive part. This scenario is well
understood even analytically and there are clear arguments which form factors have to be chosen, such that the remainder terms are negligible. If the initial interaction is purely repulsive, we are able to show that contributions of the magnetic channel induce an attractive $d$--wave pairing
interaction in the superconducting channel. 

At Van Hove filling, however, the
Fermi surface has saddle points, which cause a logarithmic
singularity in the density of states. Furthermore, near half
filling, Umklapp scattering processes become relevant. Then there is
strong mixing between particle--particle and particle--hole
channels. This case is not analytically understood yet and treated
numerically in the following. Here the form factors are chosen guided by previous $N$--patch studies and the remainder terms are neglected. A rigorous analysis of the remainder terms is left for future work. 

The numerical implementation of the boson propagator flow is eased by specifying the momentum and frequency dependence of the boson propagators. Here we first neglect the frequency dependence of the boson propagators.
Together with the frequency independent form factors, this
corresponds to a vertex function that does not depend on frequency
at all. The right hand side of the flow equations is evaluated at
boson frequency zero, which gives the main contribution. An analysis
of the two--fermion bubbles indicates that the boson propagators can
only become singular at boson momentum $(0,0)$ and $(\pi,\pi)$.
Therefore we split up each boson propagator into a part around
$(0,0)$ and a part around $(\pi,\pi)$. In these regions we
parameterize the momentum dependence of the boson propagators by
step functions, postulating a higher accuracy at the singular
points. For example, a peak of the magnetic boson propagator $M_{\mtin{S}\mtin{S}}(\mathbf{l})$ with constant form factor $f_{\mtin{S}}(\mathbf{q})=1$ at
$\mathbf{l}=(0,0)$ indicates ferromagnetic and at $\mathbf{l}=(\pi,\pi)$ antiferromagnetic
fluctuations.

In order to set up the RG flow, a regularization parameterized by an
RG scale has to be introduced. Since we are especially interested in
the interplay between $d$--wave superconductivity and
ferromagnetism, we have to choose a regularization that is sensitive
to small momentum particle--hole fluctuations \cite{TemperatureFlow}. Here we introduce a frequency
regularization by multiplying the bare propagator with the function
\begin{align}
\chi_{\Omega}(p) = \frac{p_0^2}{p_0^2 +\Omega^2} \; ,
\end{align}
where $p_0$ is the frequency part of $p$ and $\Omega$ is the RG
scale frequency. For $\Omega>0$ all graphs are regularized in the
infrared and for $\Omega\to 0$ the original model is recovered.
Unlike a Fermi surface cut--off, this regularization reproduces the
correct $(\ln \Omega)^2$--scaling of the particle--particle and $\ln
\Omega$--scaling for the particle--hole bubble at Van Hove filling.
Compared to the temperature flow, this regularization allows a clear
definition of the initial condition of the RG flow. We start the
flow at a high initial scale $\Omega_0$ and treat the scales
$\Omega>\Omega_0$ by perturbation theory in the coupling constant
$U>0$ to second order. Due to the infrared regularization,
perturbation theory converges if $U/\Omega_0$ is small enough. For large enough $\Omega_0$, we find that the dependence on $\Omega_0$ is negligible. 

{\em Results and Outlook\label{sec:results}.}
Here we present results obtained by expanding each channel with two
form factors, namely isotropic $s$--wave $f_{\mtin{S}}(\mathbf{q}) = 1$ and $d_{x^2-y^2}$--wave $f_{1}(\mathbf{q})=\cos q_1 - \cos q_2$.
Amongst others, the corresponding boson propagators allow to
detect instabilities towards $s$-- and
$d_{x^2-y^2}$--wave superconductivity, ferro-- and antiferromagnetism,
forward and exchange scattering, and also a Pomeranchuk instability.
It turned out, however, that the boson propagator flow applied to
the $(t,t')$--Hubbard model at zero temperature and Van Hove filling
already gives reasonable results if only the constant form factor is
taken into account in every channel plus the $d_{x^2-y^2}$--wave
form factor in the superconducting channel. Although $s$--wave
superconductivity is suppressed by the initial repulsion, its
screening effect is essential for the RG flow.

\begin{figure}[t]
\includegraphics[width=0.49\textwidth]{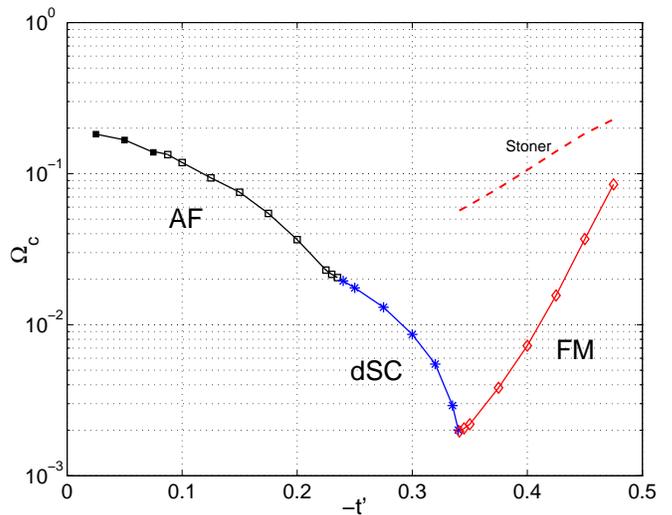}
\caption{The critical scale $\Omega_{\mbox{\tiny C}}$ in dependence
on next to nearest neighbor hopping $-t'$ at temperature zero with initial scale $\Omega_0=15t$. The 
chemical potential is set to match Van Hove filling at each $-t'$.
The instabilities of the Landau Fermi liquid are determined as
antiferromagnetism (AF), $d_{x^2-y^2}$--superconductivity (dSC), and
ferromagnetism (FM). \label{fig:OmegaKrit}}
\end{figure}

Starting from $U=3t$ we observe a
generic flow to strong coupling. If the maximum of one boson
propagator reaches the value $20t$, the flow is manually stopped at
the thus defined ''critical'' scale $\Omega_{\mbox{\tiny C}}$. Like in the previous
$N$--patch studies of the symmetric phase, this is interpreted as an
instability towards a corresponding ordered state. The critical
scale $\Omega_{\mbox{\tiny C}}$ is plotted over next to nearest
neighbor hopping $-t'/t$ in Figure \ref{fig:OmegaKrit}, where from now on $t=1$. For small
$-t'$, where Van Hove filling is close to half filling,
antiferromagnetism dominates. Since the flow is stopped at a
relatively high scale, perfect nesting--like effects are not
restricted to $-t'=0$. If $-t'$ is increased, we observe a tendency
to incommensurate antiferromagnetic order (marked with open
squares). For intermediate $-t'$ the leading instability is
$d_{x^2-y^2}$--superconductivity. It is induced by antiferromagnetic
fluctuations, as can be seen from the flow equations. For high $-t'$
ferromagnetism is the dominant instability. In the interval $-t'\in (0,0.34)$ the critical scale drops by two orders of magnitude.

The results obtained with the decomposed RG flow agree well with the temperature $N$--patch flow \cite{TemperatureFlow}. The proposed approximation to the one--loop RG flow reproduces the qualitative instabilities and the values for $-t'$, where the transitions between different instabilities take place. Indicated by the decreased critical scale, $d_{x^2-y^2}$--superconductivity and ferromagnetism suppress each other. For comparison, the Stoner criterion for ferromagnetism, which is obtained by neglecting the particle--particle channel, is plotted in Figure \ref{fig:OmegaKrit} on the ferromagnetic side.  
There is one significant deviation from the temperature flow: in our present scheme, the suppression of $\Omega_{\mtin{C}}$ in the transition region from $d$--wave superconductivity to ferromagnetism is much weaker. Since both the temperature flow and our calculation involve approxiamtions, the existence of a quantum critical point remains an open question. 

In summary, the proposed decomposition of the effective two--fermion interaction is efficient for studying competing instabilities in the $(t,t')$--Hubbard model. Separating leading from subleading processes reduces the complexity of the one--loop flow equations. Although we cannot yet rigorously justify dropping the remainder terms at Van Hove filling, there are clear arguments for this for regular Fermi surfaces. In fact, the comparison with previous $N$--patch studies shows that the qualitative structure of the RG flow is preserved. 

In the numerical implementation the momentum dependence of the boson propagators is discretized using step functions. This is more precise than the general patching in $N$--patch schemes since the choice of step functions can be guided by the one--loop bubbles. Generally, if the momentum dependence of the one--loop bubbles can be is parameterized in an analytical form, then it is possible to extract a functional parametrization of the boson propagators from the flow equations, at least for small momenta. Similarly, the dependence on small frequencies can be taken into account. Deviations from the large frequency behavior would then be subject to further remainder terms. 

The decomposition of the effective two--fermion interaction into sums of fermion bilinears interacting via boson propagators allows to decouple the fermion bilinears by multiple Hubbard Stratonovich transformations. Thereby the ambiguity of introducing boson fields is not completely removed as discussed above. However, due to the definition of the channels based on the singular momentum structure, it is reduced. Thus our results serve as an improved starting point for a continuation of the RG flow into the symmetry broken phase in a (partially) bosonized form \cite{MetznerPartial,WetterichKrahl,WetterichAFOrder,Kopietz,RGMeanField}.

\bibliography{decomposition}

\begin{thebibliography}{21}
\expandafter\ifx\csname natexlab\endcsname\relax\def\natexlab#1{#1}\fi
\expandafter\ifx\csname bibnamefont\endcsname\relax
  \def\bibnamefont#1{#1}\fi
\expandafter\ifx\csname bibfnamefont\endcsname\relax
  \def\bibfnamefont#1{#1}\fi
\expandafter\ifx\csname citenamefont\endcsname\relax
  \def\citenamefont#1{#1}\fi
\expandafter\ifx\csname url\endcsname\relax
  \def\url#1{\texttt{#1}}\fi
\expandafter\ifx\csname urlprefix\endcsname\relax\def\urlprefix{URL }\fi
\providecommand{\bibinfo}[2]{#2}
\providecommand{\eprint}[2][]{\url{#2}}

\bibitem[{\citenamefont{Zanchi and Schulz}(1998)}]{ZanchiSchulz1}
\bibinfo{author}{\bibfnamefont{D.}~\bibnamefont{Zanchi}} \bibnamefont{and}
  \bibinfo{author}{\bibfnamefont{H.~J.} \bibnamefont{Schulz}},
  \bibinfo{journal}{Europhys. Lett.} \textbf{\bibinfo{volume}{44}},
  \bibinfo{pages}{235} (\bibinfo{year}{1998}).

\bibitem[{\citenamefont{Zanchi and Schulz}(2000)}]{ZanchiSchulz2000}
\bibinfo{author}{\bibfnamefont{D.}~\bibnamefont{Zanchi}} \bibnamefont{and}
  \bibinfo{author}{\bibfnamefont{H.~J.} \bibnamefont{Schulz}},
  \bibinfo{journal}{Phys. Rev. B} \textbf{\bibinfo{volume}{61}},
  \bibinfo{pages}{13609} (\bibinfo{year}{2000}).

\bibitem[{\citenamefont{Halboth and
  Metzner}(2000{\natexlab{a}})}]{HalbothMetzner}
\bibinfo{author}{\bibfnamefont{C.~J.} \bibnamefont{Halboth}} \bibnamefont{and}
  \bibinfo{author}{\bibfnamefont{W.}~\bibnamefont{Metzner}},
  \bibinfo{journal}{Phys. Rev. B} \textbf{\bibinfo{volume}{61}},
  \bibinfo{pages}{7364} (\bibinfo{year}{2000}{\natexlab{a}}).

\bibitem[{\citenamefont{Halboth and
  Metzner}(2000{\natexlab{b}})}]{HalbothMetznerPomeranchuk}
\bibinfo{author}{\bibfnamefont{C.~J.} \bibnamefont{Halboth}} \bibnamefont{and}
  \bibinfo{author}{\bibfnamefont{W.}~\bibnamefont{Metzner}},
  \bibinfo{journal}{Phys. Rev. Lett.} \textbf{\bibinfo{volume}{85}},
  \bibinfo{pages}{5162} (\bibinfo{year}{2000}{\natexlab{b}}).

\bibitem[{\citenamefont{Zanchi}(2001)}]{ZanchiSelfEnergy}
\bibinfo{author}{\bibfnamefont{D.}~\bibnamefont{Zanchi}},
  \bibinfo{journal}{Europhys. Lett.} \textbf{\bibinfo{volume}{55}},
  \bibinfo{pages}{376} (\bibinfo{year}{2001}).

\bibitem[{\citenamefont{Honerkamp et~al.}(2001)\citenamefont{Honerkamp,
  Salmhofer, Furukawa, and Rice}}]{Umklapp}
\bibinfo{author}{\bibfnamefont{C.}~\bibnamefont{Honerkamp}},
  \bibinfo{author}{\bibfnamefont{M.}~\bibnamefont{Salmhofer}},
  \bibinfo{author}{\bibfnamefont{N.}~\bibnamefont{Furukawa}}, \bibnamefont{and}
  \bibinfo{author}{\bibfnamefont{T.~M.} \bibnamefont{Rice}},
  \bibinfo{journal}{Phys. Rev. B} \textbf{\bibinfo{volume}{63}},
  \bibinfo{pages}{035109} (\bibinfo{year}{2001}).

\bibitem[{\citenamefont{Rohe and Metzner}(2005)}]{MetznerSelfEnergy}
\bibinfo{author}{\bibfnamefont{D.}~\bibnamefont{Rohe}} \bibnamefont{and}
  \bibinfo{author}{\bibfnamefont{W.}~\bibnamefont{Metzner}},
  \bibinfo{journal}{Phys. Rev. B} \textbf{\bibinfo{volume}{71}},
  \bibinfo{pages}{115116} (\bibinfo{year}{2005}).

\bibitem[{\citenamefont{Honerkamp and Salmhofer}(2003)}]{NpatchRG}
\bibinfo{author}{\bibfnamefont{C.}~\bibnamefont{Honerkamp}} \bibnamefont{and}
  \bibinfo{author}{\bibfnamefont{M.}~\bibnamefont{Salmhofer}},
  \bibinfo{journal}{Phys. Rev. B} \textbf{\bibinfo{volume}{67}},
  \bibinfo{pages}{174504} (\bibinfo{year}{2003}).

\bibitem[{\citenamefont{Honerkamp and
  Salmhofer}(2001{\natexlab{a}})}]{TemperatureFlow}
\bibinfo{author}{\bibfnamefont{C.}~\bibnamefont{Honerkamp}} \bibnamefont{and}
  \bibinfo{author}{\bibfnamefont{M.}~\bibnamefont{Salmhofer}},
  \bibinfo{journal}{Phys. Rev. B} \textbf{\bibinfo{volume}{64}},
  \bibinfo{pages}{184516} (\bibinfo{year}{2001}{\natexlab{a}}).

\bibitem[{\citenamefont{Katanin and Kampf}(2003)}]{katanin2patch}
\bibinfo{author}{\bibfnamefont{A.~A.} \bibnamefont{Katanin}} \bibnamefont{and}
  \bibinfo{author}{\bibfnamefont{A.~P.} \bibnamefont{Kampf}},
  \bibinfo{journal}{Phys. Rev. B} \textbf{\bibinfo{volume}{68}},
  \bibinfo{pages}{195101} (\bibinfo{year}{2003}).

\bibitem[{\citenamefont{Katanin and Kampf}(2004)}]{KataninKampfSelfEnergy}
\bibinfo{author}{\bibfnamefont{A.~A.} \bibnamefont{Katanin}} \bibnamefont{and}
  \bibinfo{author}{\bibfnamefont{A.~P.} \bibnamefont{Kampf}},
  \bibinfo{journal}{Phys. Rev. Lett.} \textbf{\bibinfo{volume}{93}},
  \bibinfo{pages}{106406} (\bibinfo{year}{2004}).

\bibitem[{\citenamefont{Honerkamp et~al.}(2004)\citenamefont{Honerkamp, Rohe,
  Andergassen, and Enss}}]{gflow}
\bibinfo{author}{\bibfnamefont{C.}~\bibnamefont{Honerkamp}},
  \bibinfo{author}{\bibfnamefont{D.}~\bibnamefont{Rohe}},
  \bibinfo{author}{\bibfnamefont{S.}~\bibnamefont{Andergassen}},
  \bibnamefont{and} \bibinfo{author}{\bibfnamefont{T.}~\bibnamefont{Enss}},
  \bibinfo{journal}{Phys. Rev. B} \textbf{\bibinfo{volume}{70}},
  \bibinfo{pages}{235115} (\bibinfo{year}{2004}).

\bibitem[{\citenamefont{Honerkamp}(2001)}]{HonerkampSelfEnergy}
\bibinfo{author}{\bibfnamefont{C.}~\bibnamefont{Honerkamp}},
  \bibinfo{journal}{Eur. Phys. J. B} \textbf{\bibinfo{volume}{21}},
  \bibinfo{pages}{81} (\bibinfo{year}{2001}).

\bibitem[{\citenamefont{Honerkamp and
  Salmhofer}(2001{\natexlab{b}})}]{TemperatureFlow2}
\bibinfo{author}{\bibfnamefont{C.}~\bibnamefont{Honerkamp}} \bibnamefont{and}
  \bibinfo{author}{\bibfnamefont{M.}~\bibnamefont{Salmhofer}},
  \bibinfo{journal}{Phys. Rev. Lett.} \textbf{\bibinfo{volume}{87}},
  \bibinfo{pages}{187004} (\bibinfo{year}{2001}{\natexlab{b}}).

\bibitem[{\citenamefont{Salmhofer and Honerkamp}(2001)}]{salmfrgtt}
\bibinfo{author}{\bibfnamefont{M.}~\bibnamefont{Salmhofer}} \bibnamefont{and}
  \bibinfo{author}{\bibfnamefont{C.}~\bibnamefont{Honerkamp}},
  \bibinfo{journal}{Prog. Theor. Phys.} \textbf{\bibinfo{volume}{105}},
  \bibinfo{pages}{1} (\bibinfo{year}{2001}).

\bibitem[{\citenamefont{Karrasch et~al.}(2008)\citenamefont{Karrasch, Hedden,
  Peters, Pruschke, Sch\"onhammer, and Meden}}]{Karrasch}
\bibinfo{author}{\bibfnamefont{C.}~\bibnamefont{Karrasch}},
  \bibinfo{author}{\bibfnamefont{R.}~\bibnamefont{Hedden}},
  \bibinfo{author}{\bibfnamefont{R.}~\bibnamefont{Peters}},
  \bibinfo{author}{\bibfnamefont{T.}~\bibnamefont{Pruschke}},
  \bibinfo{author}{\bibfnamefont{K.}~\bibnamefont{Sch\"onhammer}},
  \bibnamefont{and} \bibinfo{author}{\bibfnamefont{V.}~\bibnamefont{Meden}},
  \bibinfo{journal}{J. Phys.: Cond. Matt.} \textbf{\bibinfo{volume}{20}},
  \bibinfo{pages}{345205} (\bibinfo{year}{2008}).

\bibitem[{\citenamefont{Strack et~al.}(2008)\citenamefont{Strack, Gersch, and
  Metzner}}]{MetznerPartial}
\bibinfo{author}{\bibfnamefont{P.}~\bibnamefont{Strack}},
  \bibinfo{author}{\bibfnamefont{R.}~\bibnamefont{Gersch}}, \bibnamefont{and}
  \bibinfo{author}{\bibfnamefont{W.}~\bibnamefont{Metzner}}
  (\bibinfo{year}{2008}), \eprint{arXiv:0804.3994v1}.

\bibitem[{\citenamefont{Krahl et~al.}(2008)\citenamefont{Krahl, M\"uller, and
  Wetterich}}]{WetterichKrahl}
\bibinfo{author}{\bibfnamefont{H.~C.} \bibnamefont{Krahl}},
  \bibinfo{author}{\bibfnamefont{J.~A.} \bibnamefont{M\"uller}},
  \bibnamefont{and} \bibinfo{author}{\bibfnamefont{C.}~\bibnamefont{Wetterich}}
  (\bibinfo{year}{2008}), \eprint{arXiv:0801.1773v1}.

\bibitem[{\citenamefont{Baier et~al.}(2004)\citenamefont{Baier, Bick, and
  Wetterich}}]{WetterichAFOrder}
\bibinfo{author}{\bibfnamefont{T.}~\bibnamefont{Baier}},
  \bibinfo{author}{\bibfnamefont{E.}~\bibnamefont{Bick}}, \bibnamefont{and}
  \bibinfo{author}{\bibfnamefont{C.}~\bibnamefont{Wetterich}},
  \bibinfo{journal}{Phys. Rev. B} \textbf{\bibinfo{volume}{70}},
  \bibinfo{pages}{125111} (\bibinfo{year}{2004}).

\bibitem[{\citenamefont{Sch\"utz et~al.}(2005)\citenamefont{Sch\"utz, Bartosch,
  and Kopietz}}]{Kopietz}
\bibinfo{author}{\bibfnamefont{F.}~\bibnamefont{Sch\"utz}},
  \bibinfo{author}{\bibfnamefont{L.}~\bibnamefont{Bartosch}}, \bibnamefont{and}
  \bibinfo{author}{\bibfnamefont{P.}~\bibnamefont{Kopietz}},
  \bibinfo{journal}{Phys. Rev. B} \textbf{\bibinfo{volume}{72}},
  \bibinfo{pages}{035107} (\bibinfo{year}{2005}).

\bibitem[{\citenamefont{Reiss et~al.}(2007)\citenamefont{Reiss, Rohe, and
  Metzner}}]{RGMeanField}
\bibinfo{author}{\bibfnamefont{J.}~\bibnamefont{Reiss}},
  \bibinfo{author}{\bibfnamefont{D.}~\bibnamefont{Rohe}}, \bibnamefont{and}
  \bibinfo{author}{\bibfnamefont{W.}~\bibnamefont{Metzner}},
  \bibinfo{journal}{Phys. Rev. B} \textbf{\bibinfo{volume}{75}},
  \bibinfo{pages}{075110} (\bibinfo{year}{2007}).

\end{thebibliography}
\end{document}